\documentclass[aps,prl,twocolumn,superscriptaddress,showpacs,floatfix]{revtex4}

\usepackage{bm}
\usepackage{amsmath,amssymb}
\usepackage[dvips]{graphicx,color}

\begin{document}

%\preprint{}

\title{Correlation between stress and band shift on late transition metal (111) surfaces}


\author{Yoshinori Shiihara}
\email[]{nori@telu.iis.u-tokyo.ac.jp}
%\thanks{}
\affiliation{Institute of Industrial Science, The University of Tokyo, 4-6-1 Komaba, Meguro-ku, Tokyo 153-8505, Japan}
\affiliation{Nanosystem Research Institute (NRI) "RICS", National Institute of Advanced Industrial Science and Technology (AIST), 1-1-1 Umezono, Tsukuba, Ibaraki 305-8568, Japan}

\author{Masanori Kohyama}
\affiliation{Research Institute for Ubiquitous Energy Devices (UBIQEN), AIST, 1-8-31, Midorigaoka, Ikeda, Osaka 563-8577, Japan}
\author{Shoji Ishibashi}
\affiliation{Nanosystem Research Institute (NRI) "RICS", National Institute of Advanced Industrial Science and Technology (AIST), 1-1-1 Umezono, Tsukuba, Ibaraki 305-8568, Japan}

\date{\today}

\begin{abstract}
We use \textit{ab initio} local stress calculations to investigate layer-by-layer \textit{ab initio} stress inside late transition metal (111) surfaces, focusing on the origin of stress on the surface top layer. It is found that the band shift on each surface layer is strongly correlated with the in-plane stress. For the top layer, this correlation can be explained by the Friedel model. The reduction of the local $d$-band width due to the coordination reduction is the main origin of both the $d$-band center shift and in-plane tensile stress. The changes in the directional $d$-$d$ bonding character analyzed by the in-plane and out-of-plane projected densities of states should be an additional origin of the excess tensile stress, except for Ag explained mainly by the Friedel model.
\end{abstract}

%\keywords{}

\maketitle

% References should be done using the  \cite, \ref, and \label commands
%\section{doc}
The electronic structure in a free surface is different from that in the bulk crystal. This is mainly due to the reduced coordination number (CN) of the surface atoms. This difference gives rise to the intrinsic stress on the surface because the bonding character between the surface atoms is different from that between the bulk atoms. Considering that stress can be defined as an energetic response against strain \cite{nielsen1985ssa}, we can say that the surface under stress is \textit{strained} even if the surface bond has the same bond length as that in the bulk. 

Stress on metal surfaces is an important research area in recent surface physics \cite{fiorentini1993rmf,ibach1997rss,kollar2003css,kwon2005ses,kadas2006srs,blanco2009sss,zolyomi2009srs,blanco2010ssd,punkkinen2011css}. In particular, in-plane stress and strain on 4$d$ and 5$d$ late transition metal surfaces have aroused sizable interest because of their effects on chemisorption properties in catalysis \cite{hammer2000tss,uesugi2002ies}. Mavrikakis \textit{et al.} \cite{mavrikakis1998esr} showed that increased tensile stress leads to stronger adsorption of CO or O on Pt surfaces due to a shift in the center of $d$ bands with strain. A bimetallic surface, i.e., a metal surface with an adlayer of a different metal species, is regarded as a practical instrument for producing such a strained surface \cite{hammer2000tss,chen2008mbs}. The difference in lattice constants between the adlayer and substrate metals results in stress on the adlayer and its interfaces. Because stress in atomistic scales is not easily attainable in experimental methods \cite{ibach1997rss,punkkinen2011css}, \textit{ab initio} simulation based on density-functional theory is expected to play an irreplaceable role for investigating the stress on the surface.     

In spite of the importance of stress on late transition metal surfaces, the origin of tensile stress on these surfaces is still controversial even in \textit{ab initio} theoretical arguments: in Ibach's model \cite{ibach1997rss}, the charge redistribution from a dangling bond to a surface bond strengthens the latter. K{\'o}llar \textit{et al}. \cite{kollar2003css} attributed the origin to the depletion of $sp$ electrons at the surface layer while $d$ electrons are unchanged there. Fiorentini \textit{et al}. \cite{fiorentini1993rmf} proposed a model where the depletion of antibonding $d$ electrons strengthens the surface bond. To our knowledge, these models still coexist when explaining the origin of surface stress on late transition metal surfaces \cite{blanco2009sss,punkkinen2011css}. 

In this paper, with a view to resolving this contradiction, we investigate the local stress distribution inside 4$d$ and 5$d$ late transition metal surfaces via the \textit{ab initio} local stress analysis developed in our preceding paper \cite{shiihara2010ail}, which can reveal more details about the stress state on each surface layer. We then find the correlation between the band shift and the in-plane tensile stress on each surface layer. We find that this correlation can be explained by a simple Friedel model based on the tight-binding second moment theory \cite{lannoo1991aes}: for the surface top layers of various metals, the in-plane tensile stress can be explained systematically by the reduction in the $d$-band width, inducing the $d$-band shift. Furthermore, we show that the changes in the directional $d$-$d$ bonding character observed in the in-plane and out-of-plane projected densities of states (PDOSs) can explain an additional origin of the excess tensile stress as discussed by Fiorentini \textit{et al} \cite{fiorentini1993rmf}.    

%--------------
\begin{table}
\caption{\label{a0_srelax_stress}Theoretical surface properties of 4$d$ and 5$d$ late transition metals: $a_{0}$ denotes the lattice constant; and $\gamma$ and $\tau$ are the surface energy and stress, respectively.}
%\begin{widetext}
\begin{ruledtabular}
\begin{tabular}{cccc}
   & $a_{0}$ [\AA] & $\gamma$ [J/m$^{2}$] & $\tau$ [J/m$^{2}$]\\
%    & $a_{0}$ [\AA] & $\Delta l$ [\%] & $\tau$ [J/m$^{2}$]\\
\hline
%Rh & 3.848, 3.834$^{a}$, 3.85$^{b}$ & 2.02, 2.01$^{b}$ & 2.61, 2.73$^{b}$ \\
%Pd & 3.951, 3.948$^{a}$, 3.96$^{b}$ & 1.35, 1.33$^{b}$ & 2.71, 2.57$^{b}$ \\
%Ag & 4.145, 4.152$^{a}$, 4.16$^{b}$ & 0.73, 0.76$^{b}$ & 0.82, 0.79$^{b}$ \\
%Ir & 3.883, 3.887$^{a}$, 3.877$^{c}$ & 2.31, 2.06$^{c}$ & 4.72, 4.37$^{c}$ \\
%Pt & 3.983, 3.985$^{a}$, 3.978$^{c}$ & 1.49, 1.49$^{c}$ & 4.47, 4.25$^{c}$ \\
%Au & 4.164, 4.165$^{a}$, 4.174$^{c}$ & 0.70, 0.70$^{c}$ & 2.50, 1.76$^{c}$ \\
Rh & 3.848, 3.85$^{a}$ & 2.02, 2.01$^{a}$  & 2.61, 2.73$^{a}$ \\
Pd & 3.951, 3.96$^{a}$ & 1.35, 1.33$^{a}$  & 2.71, 2.57$^{a}$ \\
Ag & 4.145, 4.16$^{a}$ & 0.73, 0.76$^{a}$  & 0.82, 0.79$^{a}$ \\
Ir & 3.883, 3.877$^{b}$ & 2.31, 2.06$^{b}$ & 4.72, 4.37$^{b}$ \\
Pt & 3.983, 3.978$^{b}$ & 1.49, 1.49$^{b}$  & 4.47, 4.25$^{b}$ \\
Au & 4.164, 4.174$^{b}$ & 0.70, 0.70$^{b}$  & 2.50, 1.76$^{b}$ \\
%Rh & 3.848, 3.85$^{a}$ & -1.77, -1.80$^{a}$   & 2.61, 2.73$^{a}$ \\
%Pd & 3.951, 3.96$^{a}$ &  0.56,  0.42$^{a}$   & 2.71, 2.57$^{a}$ \\
%Ag & 4.145, 4.16$^{a}$ & -0.16, -0.30$^{a}$   & 0.82, 0.79$^{a}$ \\
%Ir & 3.883, 3.877$^{b}$ & -2.08, -2.09$^{b}$  & 4.72, 4.37$^{b}$ \\
%Pt & 3.983, 3.978$^{b}$ & 1.17,  0.99$^{b}$   & 4.47, 4.25$^{b}$ \\
%Au & 4.164, 4.174$^{b}$ & 1.34,  1.67$^{b}$   & 2.50, 1.76$^{b}$ \\
\end{tabular}
\end{ruledtabular}
\begin{minipage}{1.0\hsize}
\begin{flushleft}
%$^{a}$ FP-LAPW, Ref.* \\
$^{a}$ PAW, Ref. \cite{kwon2005ses}\\
$^{b}$ PAW, Ref. \cite{zolyomi2009srs} \\
\end{flushleft}
\end{minipage}
\end{table}
%\end{widetext}
%------------------------
%------------------------ 
\begin{figure}
 \includegraphics[width=8.5cm,clip]{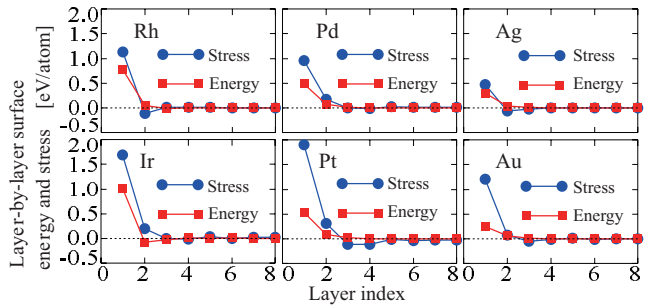}%
 \caption{In-plane layer-by-layer stress and energy distribution for a 14-layer slab of the a transition metal (111) surface. The data from the surface layer to the 8 th layer are plotted. Lines are a guide for the eyes. The in-plane stress state on an fcc (111) surface can be expressed by a single isotropic stress due to the hexagonal symmetry of the surface. }
 \label{fig_local_stress_14layer}
\end{figure} 
%------------------------
 The calculation was performed using the projector augmented wave (PAW) method \cite{blochl1994paw,kresse1999upp} with our in-house \textit{ab initio} code QMAS (Quantum MAterials Simulator) \cite{ishibashi2007aic}. We use the generalized gradient approximation \cite{perdew1996gga} for the exchange-correlation functional. The Gaussian smearing method \cite{fu1983fpc} with a smearing width of 0.1 eV is employed to introduce partial occupancy. This study deals with an fcc (111) surface which has the least surface energy among the fcc surfaces because of its high CN. The (1$\times$1) surface is simulated by a slab model where each 14-layer slab is separated by a vacuum of six-layer thickness. The \textbf{k}-point mesh in the full Brillouin zone was 32$\times$32$\times$2 for the supercells. The cutoff energy for the valence wave function was set to 544 eV in all the calculations.

The local stress is obtained by integrating the stress density \cite{filippetti2000taa} expressed in the PAW framework \cite{shiihara2010ail} over partial volumes. We can define each local region containing a single atom by setting the boundaries between atoms so that the gauge-dependent term originating from the kinetic energy density is integrated to zero in each local region. Then we obtain the local stress as the layer-by-layer stress in an fcc (111) surface slab as performed in an Al surface \cite{shiihara2010ail}. In the following, we express the layer-by-layer stress in unit of eV/atom. The sign of stress is positive for tensile stress.
    
First, we discuss the layer-by-layer stress distribution shown in Fig.\ref{fig_local_stress_14layer}. The surface lattice constants were determined so that the layer-by-layer stress is nearly zero at the center of the slab. The conventional surface energy or stress $f$ is expressed as follows:
\begin{equation}
f = \frac{1}{2S}\sum_{k}g_{k}  ,\label{stress_def}
\end{equation}
where $g$ is the layer-by-layer energy or stress in Fig.\ref{fig_local_stress_14layer}, $S$ is the unit area of the surface, and $k$ is the layer index as shown in Fig.\ref{fig_local_stress_14layer}. As shown in Table \ref{a0_srelax_stress}, the obtained surface properties are in good agreement with previous PAW calculations.

The surface stresses on late transition metal (111) surfaces are tensile. However, the stress distribution is not limited to the surface top layer but spread to the layers inside the surface as shown in Fig.\ref{fig_local_stress_14layer}. For example, in the Pt case, the third and fourth layers from the surface top layer are subjected to compressive stress. This indicates that the conventional surface stress does not directly reflect the local state of each atomic layer since it can be given as the sum of the layer-by-layer stress. Here we consider the layer-by-layer stress as a quantity expressing the local state on each atomic layer at the surface.
   
\begin{figure} 
\includegraphics[width=8.0cm,clip]{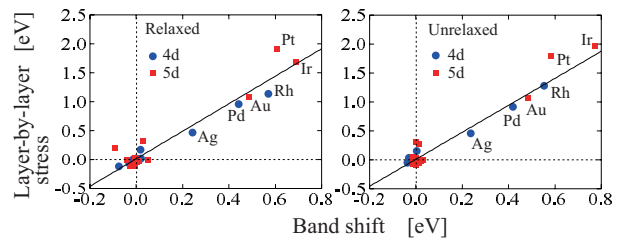}%
\caption{Correlation between the layer-by-layer in-plane stress and the shift of average band energies on relaxed and unrelaxed surfaces. The data from the surface layer to the 7 th layer are plotted. Least squares lines fitted to the data are also drawn. The labeled data are those on the surface top layers.}
\label{fig_corre_stress_bshift}
\end{figure}

We next compare the layer-by-layer stress with band shifts which are considered to directly affect the chemisorption properties of late transition metal surfaces. The band shift of each atomic layer is defined as the difference in the average band energy between the surface layer and the bulk-like layer at the center of the slab. We obtain the average band energy of each atomic layer by integrating its local density of states (LDOS). Figure \ref{fig_corre_stress_bshift} shows the results. There is a correlation between the layer-by-layer stress and the band shift on each atomic layer. For the top layers of all the late transition metal surfaces, we observe a clear correlation between the tensile stress and the upward band shift. This suggests that the origin of the stress on the surface layer can be attributed to the physics inducing the band shift. This result coincides with the fact that the tensile strain pushes up the average energy of $d$-bands on the late transition metal surface \cite{mavrikakis1998esr}. Furthermore, this figure shows that the local stress is perturbed by the surface relaxation. The effect of the surface relaxation on the surface stress is pointed out elsewhere \cite{kwon2005ses,kadas2006srs}.

%------------------------
%-------------------------
The band shift on each late transition metal surface is mainly attributed to the reduction in $d$-band width \cite{spanjaard1985scl}, because of a constant filling of the LDOS of the surface top layer to keep the charge neutrality. Thus we examine the applicability of the Friedel model, where the surface energy of a transition metal is expressed mainly by the energy increase in the surface atom caused by the reduction in the $d$-band width due to the CN reduction within the rectangular $d$-band model based on the tight-binding second moment approximation \cite{lannoo1991aes}. The in-plane surface stress can be derived from the surface energy in the Friedel model without considering any charge redistribution. By using the energy derivative with respect to strain \cite{nielsen1985ssa}, the in-plane stress $\sigma$ on the surface top layer is given as follows: 
%-----------------------
\begin{equation}
\sigma  = \frac{q}{{2Z_{\text{b}} }}\frac{{E_{\text{c}} }}
{{1 - q/p}}\left\{ {\left( {\sqrt {\frac{{Z_{\text{b}} }}
{{Z_{\text{s}} }}}  - 1} \right)\sum\limits_j {R_{j} } } \right\},\label{stress_SMA}
\end{equation}
%-----------------------
where $E_{\text{c}}$ is the bulk cohesive energy and $R_j$ is the equilibrium distance to a nearest neighboring atom $j$. The parameters of the tight-binding model, $p$ and $q$, are given in Ref. 16, and $Z_{\text{b}}$ and $Z_{\text{s}}$ are the CN of the bulk and surface atoms, respectively. Experimental values are used for $E_{\text{c}}$ and $R_j$ \cite{cleri1993}. The sum in Eq.(\ref{stress_SMA}) runs over six neighboring surface atoms on an fcc (111) surface. For convenience's sake, here we call this stress "the Friedel stress".

%------------------------
\begin{figure} 
\includegraphics[width=6.0cm,clip]{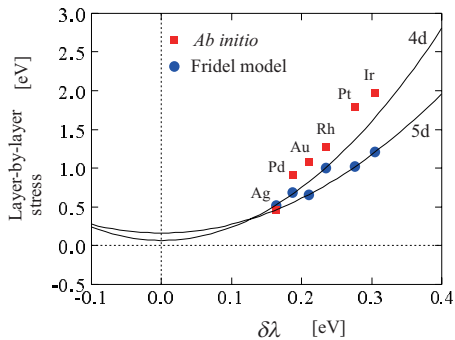}%
\caption{Comparison of the \textit{ab initio} stress on the surface top layer and the stress obtained with the Friedel model. The \textit{ab initio} stress and $\delta\lambda$ are obtained with \textit{ab initio} calculation for the unrelaxed surface. A function $\alpha\delta\lambda^2+\beta$ is fitted to 
the Friedel stress data, where $\alpha$ and $\beta$ are fitting parameters.}
\label{fig_stress_SMA}
\end{figure}
%-------------------------

Figure \ref{fig_stress_SMA} compares the Friedel stress and the \textit{ab initio} local stress. The magnitude of the Friedel stress is comparable to that of the \textit{ab initio} stress. This indicates that the Friedel model can explain the main origin of the in-plane stress on the late transition metal surface.  The agreement between the Friedel stress and the \textit{ab initio} one is impressive for Ag, implying that Ag is a typical metal whose surface properties are well described by the Friedel model.

We can assume that within the Friedel model $E_{\text{c}}$ increases quadratically with the bulk $d$-band width \cite{sutton1993electronic}. This model also shows that the band shift, the square root of the second moment of $d$-band PDOS ${\lambda}$, and its difference between the bulk and the surface $\delta\lambda = {\lambda}_\text{s}-{\lambda}_\text{b}$ are all proportional to the bulk $d$-band width \cite{lannoo1991aes}. Because the Friedel stress is governed by $E_{\text{c}}$ in Eq.(\ref{stress_SMA}), it is expected that the stress increases quadratically with the bulk $d$-band width as $E_{\text{c}}$ does. In Fig.\ref{fig_stress_SMA}, a quadratic function of $\delta\lambda$ is well fitted to the Friedel stress for each 4d or 5d surfaces because ${\delta\lambda}$ is proportional to the bulk $d$-band width in each metallic species. This explains why the \textit{ab initio} stress increases with the band shift in Fig.\ref{fig_corre_stress_bshift}, i.e., because $\delta\lambda$ can be regarded as the band shift. Such a correlation is observed only on the surface top layer in Fig.\ref{fig_stress_SMA} because the CN reduction occurs only on that layer.

\begin{figure} 
\includegraphics[width=8.5cm,clip]{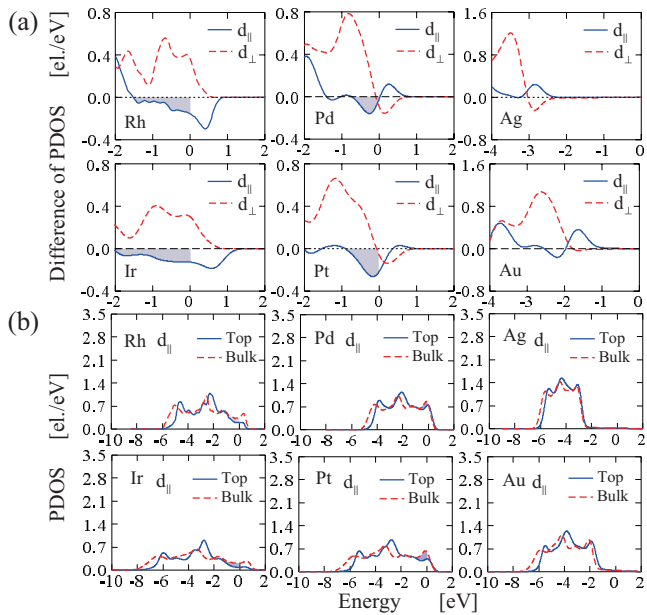}
\caption{Surface PDOSs of late transition metals: (a) difference in the surface $d_{\parallel}$- and $d_{\perp}$-band PDOS between the relaxed surface and the bulk in the vicinity of the Fermi energy, (b) whole $d_{\parallel}$ PDOS of the surface top layer and the bulk. The shaded area indicates the depleted $d$ electrons near the Fermi energy.}
\label{pdos_dperp_para}
\end{figure}

In Fig.\ref{fig_stress_SMA}, however, there exist substantial differences between the Friedel stress and the \textit{ab initio} one for the metals other than Ag, especially for 5d metals. This discrepancy is apparently caused by another mechanism that cannot be included in the Friedel model. One of the most significant effects that cannot be included in the Friedel model is the detailed changes in the PDOS in the surface atoms, other than the $d$-band width change. In particular we have to examine the changes in the directional $d$-$d$ bonding characters. Thus we have examined the in-plane and out-of-plane PDOSs of $d$ bands ($d_{\parallel}$ and $d_{\perp}$) for the surface top layers of all the metals, which were obtained as the projection to $d_{xy}$ and $d_{x^{2}-y^{2}}$, and to $d_{zx}$, $d_{yz}$, and $d_{3z^{2}-r^{2}}$, respectively (The $x$, $y$ and $z$ axes are parallel to the crystallographic directions $\left\langle {110} \right\rangle$, $\left\langle {112} \right\rangle$ and $\left\langle {111} \right\rangle$, respectively). Results are shown in Fig.\ref{pdos_dperp_para}a. We found that near the Fermi level the surface density of states is reduced in $d_{\parallel}$, while it is increased in $d_{\perp}$ compared to the bulk: the charge in the antibonding states is depleted from the in-plane surface bonds while it is not for the dangling bonds. In observing the shape of the entire PDOS in Fig.\ref{pdos_dperp_para}b \cite{comment_d_band_bottom}, we did not find any such depletion in either the bonding or nonbonding states. This directional $d$ electron depletion can cause the surface bonds to contract and result in in-plane tensile stress on the surface atoms. Furthermore, the larger depletion in Ir or Pt than in Rh or Pd (Fig.\ref{pdos_dperp_para}a) can more significantly affect the tensile stress in Ir or Pt as shown in Fig.\ref{fig_stress_SMA}. Such $d$-electron depletion is never observed for Ag because its $d$ band lies deeply below the Fermi energy. This is considered to be the reason the \textit{ab initio} stress and the Friedel one show good agreement in Ag as shown in Fig.\ref{fig_stress_SMA}. In light of the above discussion, we believe that the change in the directional $d$-$d$ bonding characters further contributes to the stress on the surface top layer on top of the contribution described by the Friedel model.  

The $d$-electron depletion as the origin of surface stress has been discussed in several studies. In late transition metal surfaces, the $d$ band shifts upward so as to keep the charge neutrality of the $d$ band as mentioned above. Fiorentini \textit{et al}. postulated that this upward shift produces depletion of $d$ electrons from the antibonding states \cite{fiorentini1993rmf}. Blanco-rey \textit{et al.} argued that the excess tensile stress comes from a decreased compressive kinetic stress component induced by directional charge redistribution from surface bonds to dangling bonds. The PDOS results we obtained regarding the antibonding $d$-electron depletion in the in-plane bonding on the surface top layer are consistent with these arguments.    

The results we obtained for Au, however, do not seem to be simply explained by the above discussion: the \textit{ab initio} stress is substantially larger than the Friedel one, while no clear antibonding $d$-electron depletion is observed in $d_{\parallel}$. This discrepancy could come from the strong $s$-$d$ hybridization, which governs the structure tendency of Au clusters with low CN \cite{hakkinen2002bonding,hasmy2008formation}. Furthermore, complicated charge redistribution in Au surfaces between $s$ and $d$ surface states has been discussed \cite{citrin1978core}. The $s$$ \leftrightarrow$$d$ redistribution can contribute to the stress on the surface top layer, especially in Au. Such relativistic effects in 5$d$ metals are not directly included in the above discussion.
%------------------------
%-------------------------

 In conclusion, we revealed that the in-plane tensile stress and the upward $d$-band shift at the top layer of late transition metal surfaces are inseparable phenomena, because both are caused by the $d$-band width reduction due to the CN reduction, as clearly described by the Friedel model. This provides us a new viewpoint on the chemical properties of late transition metal surfaces, where the aspects of both orbital interactions and structural or stress relaxation should be involved in the chemical properties. Of course, there is an additional factor that contributes to the in-plane tensile stress that is not included in the Friedel model, i.e., the changes in the directional $d$-$d$ bonding characters observed in the PDOSs of the surface top layer. The present findings should provide a new perspective for the properties of late transition metal surfaces.
 
 This work was partly supported by KAKENHI 23710107 and the Next Generation Supercomputing Project, Nanoscience Program.
%\bibliography{yshiiharaPRL20110408} 

\begin{thebibliography}{29}
\expandafter\ifx\csname natexlab\endcsname\relax\def\natexlab#1{#1}\fi
\expandafter\ifx\csname bibnamefont\endcsname\relax
  \def\bibnamefont#1{#1}\fi
\expandafter\ifx\csname bibfnamefont\endcsname\relax
  \def\bibfnamefont#1{#1}\fi
\expandafter\ifx\csname citenamefont\endcsname\relax
  \def\citenamefont#1{#1}\fi
\expandafter\ifx\csname url\endcsname\relax
  \def\url#1{\texttt{#1}}\fi
\expandafter\ifx\csname urlprefix\endcsname\relax\def\urlprefix{URL }\fi
\providecommand{\bibinfo}[2]{#2}
\providecommand{\eprint}[2][]{\url{#2}}

\bibitem[{\citenamefont{Nielsen and Martin}(1985)}]{nielsen1985ssa}
\bibinfo{author}{\bibfnamefont{O.~H.} \bibnamefont{Nielsen}} \bibnamefont{and}
  \bibinfo{author}{\bibfnamefont{R.~M.} \bibnamefont{Martin}},
  \bibinfo{journal}{Phys. Rev. B} \textbf{\bibinfo{volume}{32}},
  \bibinfo{pages}{3792} (\bibinfo{year}{1985}).

\bibitem[{\citenamefont{Fiorentini et~al.}(1993)\citenamefont{Fiorentini,
  Methfessel, and Scheffler}}]{fiorentini1993rmf}
\bibinfo{author}{\bibfnamefont{V.}~\bibnamefont{Fiorentini}},
  \bibinfo{author}{\bibfnamefont{M.}~\bibnamefont{Methfessel}},
  \bibnamefont{and}
  \bibinfo{author}{\bibfnamefont{M.}~\bibnamefont{Scheffler}},
  \bibinfo{journal}{Phys. Rev. Lett.} \textbf{\bibinfo{volume}{71}},
  \bibinfo{pages}{1051} (\bibinfo{year}{1993}).

\bibitem[{\citenamefont{Ibach}(1997)}]{ibach1997rss}
\bibinfo{author}{\bibfnamefont{H.}~\bibnamefont{Ibach}},
  \bibinfo{journal}{Surf. sci. rep.} \textbf{\bibinfo{volume}{29}},
  \bibinfo{pages}{193} (\bibinfo{year}{1997}).

\bibitem[{\citenamefont{Koll{\'a}r et~al.}(2003)\citenamefont{Koll{\'a}r,
  Vitos, Osorio-Guillen, and Ahuja}}]{kollar2003css}
\bibinfo{author}{\bibfnamefont{J.}~\bibnamefont{Koll{\'a}r}},
  \bibinfo{author}{\bibfnamefont{L.}~\bibnamefont{Vitos}},
  \bibinfo{author}{\bibfnamefont{J.~M.} \bibnamefont{Osorio-Guillen}},
  \bibnamefont{and} \bibinfo{author}{\bibfnamefont{R.}~\bibnamefont{Ahuja}},
  \bibinfo{journal}{Phys. Rev. B} \textbf{\bibinfo{volume}{68}},
  \bibinfo{pages}{245417} (\bibinfo{year}{2003}).

\bibitem[{\citenamefont{Kwon et~al.}(2005)\citenamefont{Kwon, Nabi, K{\'a}das,
  Vitos, Koll{\'a}r, Johansson, and Ahuja}}]{kwon2005ses}
\bibinfo{author}{\bibfnamefont{S.~K.} \bibnamefont{Kwon}},
  \bibinfo{author}{\bibfnamefont{Z.}~\bibnamefont{Nabi}},
  \bibinfo{author}{\bibfnamefont{K.}~\bibnamefont{K{\'a}das}},
  \bibinfo{author}{\bibfnamefont{L.}~\bibnamefont{Vitos}},
  \bibinfo{author}{\bibfnamefont{J.}~\bibnamefont{Koll{\'a}r}},
  \bibinfo{author}{\bibfnamefont{B.}~\bibnamefont{Johansson}},
  \bibnamefont{and} \bibinfo{author}{\bibfnamefont{R.}~\bibnamefont{Ahuja}},
  \bibinfo{journal}{Phys. Rev. B} \textbf{\bibinfo{volume}{72}},
  \bibinfo{pages}{235423} (\bibinfo{year}{2005}).

\bibitem[{\citenamefont{K{\'a}das et~al.}(2006)\citenamefont{K{\'a}das, Nabi,
  Kwon, Vitos, Ahuja, Johansson, and Koll{\'a}r}}]{kadas2006srs}
\bibinfo{author}{\bibfnamefont{K.}~\bibnamefont{K{\'a}das}},
  \bibinfo{author}{\bibfnamefont{Z.}~\bibnamefont{Nabi}},
  \bibinfo{author}{\bibfnamefont{S.~K.} \bibnamefont{Kwon}},
  \bibinfo{author}{\bibfnamefont{L.}~\bibnamefont{Vitos}},
  \bibinfo{author}{\bibfnamefont{R.}~\bibnamefont{Ahuja}},
  \bibinfo{author}{\bibfnamefont{B.}~\bibnamefont{Johansson}},
  \bibnamefont{and}
  \bibinfo{author}{\bibfnamefont{J.}~\bibnamefont{Koll{\'a}r}},
  \bibinfo{journal}{Surf. Sci.} \textbf{\bibinfo{volume}{600}},
  \bibinfo{pages}{395} (\bibinfo{year}{2006}).

\bibitem[{\citenamefont{Blanco-Rey et~al.}(2009)\citenamefont{Blanco-Rey,
  Pratt, and Jenkins}}]{blanco2009sss}
\bibinfo{author}{\bibfnamefont{M.}~\bibnamefont{Blanco-Rey}},
  \bibinfo{author}{\bibfnamefont{S.~J.} \bibnamefont{Pratt}}, \bibnamefont{and}
  \bibinfo{author}{\bibfnamefont{S.~J.} \bibnamefont{Jenkins}},
  \bibinfo{journal}{Phys. Rev. Lett.} \textbf{\bibinfo{volume}{102}},
  \bibinfo{pages}{026102} (\bibinfo{year}{2009}).

\bibitem[{\citenamefont{Z{\'o}lyomi et~al.}(2009)\citenamefont{Z{\'o}lyomi,
  Vitos, Kwon, and Koll{\'a}r}}]{zolyomi2009srs}
\bibinfo{author}{\bibfnamefont{V.}~\bibnamefont{Z{\'o}lyomi}},
  \bibinfo{author}{\bibfnamefont{L.}~\bibnamefont{Vitos}},
  \bibinfo{author}{\bibfnamefont{S.~K.} \bibnamefont{Kwon}}, \bibnamefont{and}
  \bibinfo{author}{\bibfnamefont{J.}~\bibnamefont{Koll{\'a}r}},
  \bibinfo{journal}{J. Phys.: Cond. Matt.} \textbf{\bibinfo{volume}{21}},
  \bibinfo{pages}{095007} (\bibinfo{year}{2009}).

\bibitem[{\citenamefont{Blanco-Rey and Jenkins}(2010)}]{blanco2010ssd}
\bibinfo{author}{\bibfnamefont{M.}~\bibnamefont{Blanco-Rey}} \bibnamefont{and}
  \bibinfo{author}{\bibfnamefont{S.~J.} \bibnamefont{Jenkins}},
  \bibinfo{journal}{J. Phys.: Cond. Matt.} \textbf{\bibinfo{volume}{22}},
  \bibinfo{pages}{135007} (\bibinfo{year}{2010}).

\bibitem[{\citenamefont{Punkkinen et~al.}(2011)\citenamefont{Punkkinen, Kwon,
  Koll{\'a}r, Johansson, and Vitos}}]{punkkinen2011css}
\bibinfo{author}{\bibfnamefont{M.~P.~J.} \bibnamefont{Punkkinen}},
  \bibinfo{author}{\bibfnamefont{S.~K.} \bibnamefont{Kwon}},
  \bibinfo{author}{\bibfnamefont{J.}~\bibnamefont{Koll{\'a}r}},
  \bibinfo{author}{\bibfnamefont{B.}~\bibnamefont{Johansson}},
  \bibnamefont{and} \bibinfo{author}{\bibfnamefont{L.}~\bibnamefont{Vitos}},
  \bibinfo{journal}{Phys. Rev. Lett.} \textbf{\bibinfo{volume}{106}},
  \bibinfo{pages}{057202} (\bibinfo{year}{2011}).

\bibitem[{\citenamefont{Hammer and N{\o}rskov}(2000)}]{hammer2000tss}
\bibinfo{author}{\bibfnamefont{B.}~\bibnamefont{Hammer}} \bibnamefont{and}
  \bibinfo{author}{\bibfnamefont{J.~K.} \bibnamefont{N{\o}rskov}},
  \bibinfo{journal}{Adv. Catal.} \textbf{\bibinfo{volume}{45}},
  \bibinfo{pages}{71} (\bibinfo{year}{2000}).

\bibitem[{\citenamefont{Uesugi-Saitow and Yata}(2002)}]{uesugi2002ies}
\bibinfo{author}{\bibfnamefont{Y.}~\bibnamefont{Uesugi-Saitow}}
  \bibnamefont{and} \bibinfo{author}{\bibfnamefont{M.}~\bibnamefont{Yata}},
  \bibinfo{journal}{Phys. Rev. Lett.} \textbf{\bibinfo{volume}{88}},
  \bibinfo{pages}{256104} (\bibinfo{year}{2002}).

\bibitem[{\citenamefont{Mavrikakis et~al.}(1998)\citenamefont{Mavrikakis,
  Hammer, and N{\o}rskov}}]{mavrikakis1998esr}
\bibinfo{author}{\bibfnamefont{M.}~\bibnamefont{Mavrikakis}},
  \bibinfo{author}{\bibfnamefont{B.}~\bibnamefont{Hammer}}, \bibnamefont{and}
  \bibinfo{author}{\bibfnamefont{J.~K.} \bibnamefont{N{\o}rskov}},
  \bibinfo{journal}{Phys. Rev. Lett.} \textbf{\bibinfo{volume}{81}},
  \bibinfo{pages}{2819} (\bibinfo{year}{1998}).

\bibitem[{\citenamefont{Chen et~al.}(2008)\citenamefont{Chen, Menning, and
  Zellner}}]{chen2008mbs}
\bibinfo{author}{\bibfnamefont{J.~G.} \bibnamefont{Chen}},
  \bibinfo{author}{\bibfnamefont{C.~A.} \bibnamefont{Menning}},
  \bibnamefont{and} \bibinfo{author}{\bibfnamefont{M.~B.}
  \bibnamefont{Zellner}}, \bibinfo{journal}{Surf. Sci. Rep.}
  \textbf{\bibinfo{volume}{63}}, \bibinfo{pages}{201} (\bibinfo{year}{2008}).

\bibitem[{\citenamefont{Shiihara et~al.}(2010)\citenamefont{Shiihara, Kohyama,
  and Ishibashi}}]{shiihara2010ail}
\bibinfo{author}{\bibfnamefont{Y.}~\bibnamefont{Shiihara}},
  \bibinfo{author}{\bibfnamefont{M.}~\bibnamefont{Kohyama}}, \bibnamefont{and}
  \bibinfo{author}{\bibfnamefont{S.}~\bibnamefont{Ishibashi}},
  \bibinfo{journal}{Phys. Rev. B} \textbf{\bibinfo{volume}{81}},
  \bibinfo{pages}{075441} (\bibinfo{year}{2010}).

\bibitem[{\citenamefont{Lannoo and Friedel}(1991)}]{lannoo1991aes}
\bibinfo{author}{\bibfnamefont{M.}~\bibnamefont{Lannoo}} \bibnamefont{and}
  \bibinfo{author}{\bibfnamefont{P.}~\bibnamefont{Friedel}},
  \emph{\bibinfo{title}{{Atomic and electronic structure of surfaces:
  theoretical foundations}}} (\bibinfo{publisher}{Springer-Verlag},
  \bibinfo{year}{1991}).

\bibitem[{\citenamefont{Bl{\"o}chl}(1994)}]{blochl1994paw}
\bibinfo{author}{\bibfnamefont{P.~E.} \bibnamefont{Bl{\"o}chl}},
  \bibinfo{journal}{Phys. Rev. B} \textbf{\bibinfo{volume}{50}},
  \bibinfo{pages}{17953} (\bibinfo{year}{1994}).

\bibitem[{\citenamefont{Kresse and Joubert}(1999)}]{kresse1999upp}
\bibinfo{author}{\bibfnamefont{G.}~\bibnamefont{Kresse}} \bibnamefont{and}
  \bibinfo{author}{\bibfnamefont{D.}~\bibnamefont{Joubert}},
  \bibinfo{journal}{Phys. Rev. B} \textbf{\bibinfo{volume}{59}},
  \bibinfo{pages}{1758} (\bibinfo{year}{1999}).

\bibitem[{\citenamefont{Ishibashi et~al.}(2007)\citenamefont{Ishibashi, Tamura,
  Tanaka, Kohyama, and Terakura}}]{ishibashi2007aic}
\bibinfo{author}{\bibfnamefont{S.}~\bibnamefont{Ishibashi}},
  \bibinfo{author}{\bibfnamefont{T.}~\bibnamefont{Tamura}},
  \bibinfo{author}{\bibfnamefont{S.}~\bibnamefont{Tanaka}},
  \bibinfo{author}{\bibfnamefont{M.}~\bibnamefont{Kohyama}}, \bibnamefont{and}
  \bibinfo{author}{\bibfnamefont{K.}~\bibnamefont{Terakura}},
  \bibinfo{journal}{Phys. Rev. B} \textbf{\bibinfo{volume}{76}},
  \bibinfo{pages}{153310} (\bibinfo{year}{2007}).

\bibitem[{\citenamefont{Perdew et~al.}(1996)\citenamefont{Perdew, Burke, and
  Ernzerhof}}]{perdew1996gga}
\bibinfo{author}{\bibfnamefont{J.~P.} \bibnamefont{Perdew}},
  \bibinfo{author}{\bibfnamefont{K.}~\bibnamefont{Burke}}, \bibnamefont{and}
  \bibinfo{author}{\bibfnamefont{M.}~\bibnamefont{Ernzerhof}},
  \bibinfo{journal}{Phys. Rev. Lett.} \textbf{\bibinfo{volume}{77}},
  \bibinfo{pages}{3865} (\bibinfo{year}{1996}).

\bibitem[{\citenamefont{Fu and Ho}(1983)}]{fu1983fpc}
\bibinfo{author}{\bibfnamefont{C.~L.} \bibnamefont{Fu}} \bibnamefont{and}
  \bibinfo{author}{\bibfnamefont{K.~M.} \bibnamefont{Ho}},
  \bibinfo{journal}{Phys. Rev. B} \textbf{\bibinfo{volume}{28}},
  \bibinfo{pages}{5480} (\bibinfo{year}{1983}).

\bibitem[{\citenamefont{Filippetti and Fiorentini}(2000)}]{filippetti2000taa}
\bibinfo{author}{\bibfnamefont{A.}~\bibnamefont{Filippetti}} \bibnamefont{and}
  \bibinfo{author}{\bibfnamefont{V.}~\bibnamefont{Fiorentini}},
  \bibinfo{journal}{Phys. Rev. B} \textbf{\bibinfo{volume}{61}},
  \bibinfo{pages}{8433} (\bibinfo{year}{2000}).

\bibitem[{\citenamefont{Spanjaard et~al.}(1985)\citenamefont{Spanjaard,
  Guillot, Desjonqu{\`e}res, Tr{\'e}glia, and Lecante}}]{spanjaard1985scl}
\bibinfo{author}{\bibfnamefont{D.}~\bibnamefont{Spanjaard}},
  \bibinfo{author}{\bibfnamefont{C.}~\bibnamefont{Guillot}},
  \bibinfo{author}{\bibfnamefont{M.~C.} \bibnamefont{Desjonqu{\`e}res}},
  \bibinfo{author}{\bibfnamefont{G.}~\bibnamefont{Tr{\'e}glia}},
  \bibnamefont{and} \bibinfo{author}{\bibfnamefont{J.}~\bibnamefont{Lecante}},
  \bibinfo{journal}{Surf. Sci. Rep.} \textbf{\bibinfo{volume}{5}},
  \bibinfo{pages}{1} (\bibinfo{year}{1985}).

\bibitem[{\citenamefont{Cleri and Rosato}(1993)}]{cleri1993}
\bibinfo{author}{\bibfnamefont{F.}~\bibnamefont{Cleri}} \bibnamefont{and}
  \bibinfo{author}{\bibfnamefont{V.}~\bibnamefont{Rosato}},
  \bibinfo{journal}{Phys. Rev. B} \textbf{\bibinfo{volume}{48}},
  \bibinfo{pages}{22} (\bibinfo{year}{1993}).

\bibitem[{\citenamefont{Sutton}(1993)}]{sutton1993electronic}
\bibinfo{author}{\bibfnamefont{A.~P.} \bibnamefont{Sutton}},
  \emph{\bibinfo{title}{Electronic structure of materials}}
  (\bibinfo{publisher}{Oxford University Press, USA}, \bibinfo{year}{1993}).

\bibitem[{com()}]{comment_d_band_bottom}
\bibinfo{note}{Note that the bonding peak at the $d$-band bottom is enhanced in
  each $d_{\parallel}$, while it is shifted upward in the energy due to the
  band-width reduction.}

\bibitem[{\citenamefont{H{\"a}kkinen et~al.}(2002)\citenamefont{H{\"a}kkinen,
  Moseler, and Landman}}]{hakkinen2002bonding}
\bibinfo{author}{\bibfnamefont{H.}~\bibnamefont{H{\"a}kkinen}},
  \bibinfo{author}{\bibfnamefont{M.}~\bibnamefont{Moseler}}, \bibnamefont{and}
  \bibinfo{author}{\bibfnamefont{U.}~\bibnamefont{Landman}},
  \bibinfo{journal}{Phys. Rev. Lett.} \textbf{\bibinfo{volume}{89}},
  \bibinfo{pages}{033401} (\bibinfo{year}{2002}).

\bibitem[{\citenamefont{Hasmy et~al.}(2008)\citenamefont{Hasmy, Rinc{\'o}n,
  Hern{\'a}ndez, Mujica, M{\'a}rquez, and Gonz{\'a}lez}}]{hasmy2008formation}
\bibinfo{author}{\bibfnamefont{A.}~\bibnamefont{Hasmy}},
  \bibinfo{author}{\bibfnamefont{L.}~\bibnamefont{Rinc{\'o}n}},
  \bibinfo{author}{\bibfnamefont{R.}~\bibnamefont{Hern{\'a}ndez}},
  \bibinfo{author}{\bibfnamefont{V.}~\bibnamefont{Mujica}},
  \bibinfo{author}{\bibfnamefont{M.}~\bibnamefont{M{\'a}rquez}},
  \bibnamefont{and}
  \bibinfo{author}{\bibfnamefont{C.}~\bibnamefont{Gonz{\'a}lez}},
  \bibinfo{journal}{Phys. Rev. B} \textbf{\bibinfo{volume}{78}},
  \bibinfo{pages}{115409} (\bibinfo{year}{2008}).

\bibitem[{\citenamefont{Citrin et~al.}(1978)\citenamefont{Citrin, Wertheim, and
  Baer}}]{citrin1978core}
\bibinfo{author}{\bibfnamefont{P.~H.} \bibnamefont{Citrin}},
  \bibinfo{author}{\bibfnamefont{G.~K.} \bibnamefont{Wertheim}},
  \bibnamefont{and} \bibinfo{author}{\bibfnamefont{Y.}~\bibnamefont{Baer}},
  \bibinfo{journal}{Phys. Rev. Lett.} \textbf{\bibinfo{volume}{41}},
  \bibinfo{pages}{1425} (\bibinfo{year}{1978}).

\end{thebibliography}

\end{document}